\documentstyle[aps,epsf]{revtex}  
\begin{document}        

\baselineskip 14pt
\title{$CP$ Violation in $B$ Decays in a 2-Higgs Doublet Model 
for the Top Quark\footnote{
Talk presented by G.-H. Wu at DPF'99, Los Angeles, California, 
5-9 January 1999.}}

\author{Ken Kiers}
\address{Physics Department, Taylor University, 236 West Reade Ave., Upland,
IN 46989, USA}
\author{Amarjit Soni}
\address{High Energy Theory, Department of Physics,
Brookhaven National Laboratory,Upton, NY 11973-5000, USA}
\author{Guo-Hong Wu}
\address{Department of Physics, Purdue University,
        West Lafayette, IN 47907, USA}
\maketitle              

\begin{abstract}     
In the absence of natural flavor conservation, multi-Higgs doublet
models generally contain new sources of $CP$ violation and anomalous
charged Higgs Yukawa couplings.
 We present a charged-Higgs $CP$ violation study of one 
such two-Higgs doublet model
(2HDM) which treats the top quark differently from the other quarks.
The phenomenological implications  for the $K\bar{K}$ system and 
for $B$ decays
differ significantly from those of the standard model (SM) and of the 2HDM's
with natural flavor conservation.
In particular, the SM phase in this model could take a wide range 
of values, and the $CP$ asymmetry in the ``gold-plated" decay mode
$B \to J/\psi K_S$ could be quite different from 
and even of opposite sign relative to the SM prediction.
A new mechanism is also noted for generating a large neutron electric 
dipole moment which is close to the present experimental limit. 
\end{abstract}   

\section{Introduction}               

  Though $CP$ violation has only been observed in the neutral kaon system,
both the standard model (SM) and its many extensions predict large
$CP$-violating effects in $B$ decays. 
One of the main programs at the $B$ factories is to test the
SM Cabibbo-Kobayashi-Maskawa (CKM) paradigm of $CP$ violation
through, for example, measurement of the angles of unitarity
triangle. In particular, the angle 
$\beta_{\rm CKM} \equiv \arg\left(-V_{cd}V_{cb}^\ast/
V_{td}V_{tb}^\ast\right)$ can be cleanly related within the SM
 to the $CP$ asymmetry
in the ``gold-plated" decay of $B \to J/\psi K_S$.
The current SM fit already places quite a nontrivial  constraint on
this angle: $\sin(2\beta_{\rm CKM}) = .75\pm.10$ \cite{paganini}.
Therefore direct experimental measurement of the angles of the
unitarity triangle 
with values in gross disagreement with the SM expectation will be
a clear signal for new sources of $CP$ violation.
It is thus important to explore the pattern
of $CP$ violation in $B$ decays in  different models of $CP$ violation. 

 Two Higgs doublet models (2HDM) without natural flavor conservation
\cite{nfc,nonfc} (NFC),
also called Model III \cite{type3}, 
contain in general both tree-level flavor-changing neutral Higgs (FCNH) 
couplings and new $CP$-violating phases beside 
the CKM phase. The phenomenology of this class of models thus differs
significantly from that of the much studied 2HDM's with NFC 
({\it i.e.,} Models I and II \cite{hhunter}) where neither tree level 
FCNH couplings nor new $CP$ violating phases
are present.  It is worth noting that the abandonment of NFC is not 
against any fundamental principle and in fact it opens up many possibilities
for flavor physics. Depending on the ansatz for FCNH couplings,
type III 2HDM need not run into problems with the low energy data
such as $K\bar{K}$ mixing.
  In this talk, we present a $CP$ violation study of a specific
type III 2HDM introduced in \cite{daskao} where the top quark is treated
differently than the other quarks. As will be shown, the charged
Higgs sector of the model could lead to a distinctive pattern
of $CP$ violation in $B$ decays 
even after imposing all the experimental constraints.

\section{The T2HDM}

  The top-quark 2HDM (T2HDM)~\cite{daskao} distinguishes itself from 2HDM's with
NFC by the special status it gives to the top quark, as evidenced
in the Yukawa couplings,
\begin{equation}
        {\cal L}_Y  =  -\overline{L}_L\phi_1 E\ell_R
                -\overline{Q}_L\phi_1 Fd_R
                -\overline{Q}_L\widetilde{\phi}_1 G {\bf 1^{(1)}} u_R
                -\overline{Q}_L\widetilde{\phi}_2 G {\bf 1^{(2)}} u_R
                + {\rm h.c.},
        \label{eq:yuk}
\end{equation}
\noindent where $\widetilde{\phi}_i = i\sigma^2\phi_i^\ast$ $(i=1,2)$,
and where $E$, $F$ and $G$ are $3\times3$ matrices in generation space;
${\bf 1^{(1)}}\equiv {\rm diag}(1,1,0)$;
${\bf 1^{(2)}}\equiv {\rm diag}(0,0,1)$;
and $Q_L$ and $L_L$ are the usual left-handed quark and lepton doublets.
Note that among the right-handed quarks, only $t_R$ couples to 
$\phi_2$, and that the mass hierarchy between the top quark and the other
quarks is understood as a result of the hierarchy between the VEV's of 
$\phi_2$ and $\phi_1$.
This form of the Yukawa interactions can be seen as a consequence of some
discrete symmetry.
Denoting the ratio of the two VEV's by 
$\tan\beta=v_2/v_1$, the model then requires a large $\tan\beta$
in accordance with the large mass ratio of the top and bottom quarks,
 $m_t/m_b$. In our analysis, we will always take $\tan \beta \ge 10$.

  The absence of NFC in the model results in tree level FCNH couplings
among the up-type quarks, whose contribution to $D\bar{D}$ mixing
is dependent on the mixing among the right-handed up-type quarks\cite{daskao}.
For the charged Higgs sector, we obtain the following Yukawa
Lagrangian \cite{ksw},
\begin{eqnarray}
        {\cal L}^C_Y & = & (g/\sqrt{2}m_W)\left\{
                -\overline{u}_L V m_d d_R\left[G^+ -\tan\beta H^+\right]
                +\overline{u}_R m_u V d_L\left[G^+ -\tan\beta H^+\right]
                        \right. \nonumber \\ 
                & & \;\;\;\;\;\;\;\;\;\;\;\;\;\; \left.
                +\overline{u}_R \Sigma^\dagger V d_L\left[\tan\beta +
                        \cot\beta\right]H^+ +{\rm h.c.}\right\},
        \label{eq:chiggs}
\end{eqnarray}
\noindent where $G^\pm$ and $H^\pm$ represent the would-be Goldstone bosons
and the physical charged Higgs bosons, respectively.
Here $m_u$ and $m_d$ are the diagonal up- and down-type quark mass matrices,
$V$ is the usual CKM matrix, and
$\Sigma \equiv m_u U_R^\dagger {\bf 1^{(2)}} U_R$ with $U_R$ being
the unitary rotation that brings the right-handed up-type quarks from gauge
eigenstates to mass eigenstates.
Unlike 2HDM's with NFC, the T2HDM depends on the $U_R$ rotation.
For our analysis, we take the simple parameterization for $U_R$ 
\cite{daskao,ksw},
\begin{equation}
U_R = \left( \begin{array}{ccc}
\cos\phi & -\sin\phi & 0 \\
\sin\phi & \cos\phi & 0 \\
0 & 0 & 1
\end{array} \right)
\left( \begin{array}{ccc}
1 & 0 & 0 \\
0 & \sqrt{1-|\epsilon_{ct}\xi|^2} & -\epsilon_{ct}\xi^\ast \\
0 & \epsilon_{ct}\xi & \sqrt{1-|\epsilon_{ct}\xi|^2}
\end{array} \right),
\end{equation}
\noindent where $\epsilon_{ct}\equiv m_c/m_t$ and 
$\xi=|\xi| e^{-i\delta}$ is a complex number of order unity. 
The $\Sigma$ matrix then depends only on the unknown parameter
$\xi$,
\begin{equation}
        \Sigma = \left(\begin{array}{ccc}
         0 & 0 & 0 \\
         0 & m_c \epsilon_{ct}^2|\xi|^2 &
         m_c\epsilon_{ct}\xi^\ast\sqrt{1-|\epsilon_{ct}\xi|^2} \\
         0 & m_c\xi\sqrt{1-|\epsilon_{ct}\xi|^2} &
         m_t\left(1-|\epsilon_{ct}\xi|^2\right)
                \end{array} \right) .
\end{equation}
For numerical analysis, we will assume $|\xi|=1$.

  Several features of the model can be noted.
First, it contains a new CP violating phase $\delta$ through its
dependence on $U_R$. By comparison, 2HDM's with NFC (Models I and II)
involve only the CKM phase.
Secondly, charm quark Yukawa couplings of the type $H^+\bar{c}_Rq_L$
($q=d,s,b$) are proportional to $\tan\beta$ and thus are enhanced
in this model. In contrast, the corresponding couplings in Models I and II
of the 2HDM are suppressed by $1/\tan\beta$.
It is the presence of the new phase that leads to a novel CP-violating 
pattern in $B$ decays, as will be  shown later. 
On the other hand, the enhanced charm Yukawa's draw our immediate attention
to the $K\bar{K}$ system, $b\to s \gamma$ and $B \to X_c \tau \nu$.

\section{Experimental Constraints}
\subsection{the $K\bar{K}$ system}

 Recall that  there exists no tree level FCNH coupling among the down-type
quarks in the present model.
  The most important contribution to the $K\bar{K}$ system from the charged
Higgs sector is associated with the $HHcc$ box diagram which has a huge
enhancement factor of $\tan^4 \beta$.  
Therefore, both the $K\bar{K}$ mass difference $\Delta m_K$ and 
the CP violation parameter $\epsilon_K$ are expected to place
stringent constraints on the parameter space of the model.

{\bf $1. \;\; \Delta m_K$}

 In the SM the short-distance contribution to $\Delta m_K$
is dominated by the charm quark and the long distance contribution
can not be estimated reliably though is expected to be quite significant
\cite{buchalla}.
Including both the SM and the charged Higgs effects, we obtain in the T2HDM
the total short distance contribution to the mass difference as,
\begin{equation}
(\Delta m_K)_{SD}  = (G_F^2/6\pi^2) f_K^2 B_K m_K \lambda_c^2
\times \left[ m_c^2 \eta_1 + (m_c^4 \tan^4 \beta/4 m_H^2) \eta^{\prime}_1
                \right]
\end{equation}
\noindent where the first term is the SM charm contribution and the second
term is due to the dominant $HHcc$ box diagram.
Here $B_K$ is the usual bag factor, 
$\lambda_c=V_{cs}V_{cd}^*$, and $\eta_1$ and
$\eta^{\prime}_1$ are the QCD corrections to the two box diagrams.
The SM top quark contribution is a few percent of the charm quark
effect and is not included here. Similarly, in comparison to the 
$HHcc$ box diagram,
the contributions from $WHcc$ and other box diagrams are small
in the large $\tan \beta$ limit and are ignored.

  To numerically deduce the allowed parameter space subject to
the $\Delta m_K$ constraint, we use the method
described in \cite{paganini} for error analysis.
 Assuming the magnitude of the long distance
contribution to $\Delta m_K$ to be no larger than $30\%$ and simply taking
$\eta_1^{\prime}=\eta_1$, we find
the $95\%$ C.L. limit,
\begin{equation}
 m_H/\tan^2 \beta  >  0.48 \;{\rm GeV} \,
\end{equation}
which is valid for $\tan \beta > 10$. 
As a result of the $\tan^4 \beta$ dependence in $\Delta m_K$,
 this constraint puts a severe lower bound on the charged Higgs mass
when $\tan \beta$ is large.
By comparison, the charged Higgs effect on $\Delta m_K$ in 2HDM's with NFC 
is suppressed by $\cot^4\beta$ and is thus negligible.
Note also that the short distance effect 
is independent of the mixing parameter $\xi$ of the T2HDM.

{\bf $2. \;\; \epsilon_K$}

  Unlike $\Delta m_K$, the $CP$ violation parameter $\epsilon_K$ 
is short distance dominated.  In the SM, the dominant component
of $\epsilon_K$ comes from the $tt$ and $tc$ box diagrams. The complete
expression and its numerical evaluation, including 
the next-to-leading-order (NLO) QCD corrections, can be found in 
\cite{buchalla}.
The charged Higgs contribution to $\epsilon_K$ is still dominated by the
charm quark, with the imaginary part of the  $HHcc$ box diagram now
dependent on both the magnitude and the phase of $\xi$.
More explicitly, this contribution from charged Higgs exchange
is given by, 
\begin{equation} \label{eq:epsK}
\epsilon_K^H  =  e^{i
\frac{\pi}{4}} C_{\epsilon} B_K A \lambda^4
 \eta_1^{\prime} \sqrt{\rho^2 + \eta^2} \sin(\gamma + \delta) |\xi|
  (m_c \tan \beta)^4/4m_W^2 m^2_H
\end{equation}
\noindent where $A$, $\lambda$, $\rho$, and $\eta$ are the CKM parameters
in the Wolfenstein parameterization \cite{wolfen},
$\gamma \equiv \tan^{-1} \eta/\rho$ is the CKM phase, and
$C_{\epsilon}=G^2_F f^2_K m^2_W m_K/6 \sqrt{2} \pi^2 \Delta m_K
=3.78 \times 10^4$.

   Due to its dependence on the new phase $\delta$, $\epsilon_K$ 
no longer restricts the CKM angle  $\gamma$. In fact, current
data allows a wide range of values for $\gamma$.
We can obtain bounds on the parameter
$Y \equiv  \sin (\gamma+\delta) |\xi| 
( \tan \beta/20)^4( 200 \; \mbox{GeV}/m_H)^2$
for any given value of $\gamma$
by allowing $\sqrt{\rho^2 + \eta^2}$ to vary within its $1\sigma$
uncertainties derived from $b\rightarrow u e \nu$.  
For a real CKM matrix ($\gamma=0^\circ$), we obtain the $95\%$ C.L bound 
$0.08 <  Y < 0.39$; the bound becomes
$0.14 <  Y < 0.65$ for $\gamma=-45^\circ$. 
  If $\gamma$ takes its SM central value of $68^\circ$
\cite{paganini}, the bound becomes nearly symmetric about zero as
expected: $-0.085 < Y  < 0.08$. 
Unlike  $\Delta m_K$, the CP violation parameter 
$\epsilon_K$ imposes a constraint on the $(m_H, \tan \beta)$ plane
that is dependent on $|\xi|$ and $\delta$.
As for $\Delta m_K$, the charged Higgs contribution to $\epsilon_K$ in
2HDM's with NFC is suppressed by $\cot^4\beta$ and is thus negligible
when $\tan\beta$ is large.

\subsection{the decay $b\to s \gamma$}

  The agreement between the experimental measurement and the SM prediction 
for the $b\to s \gamma$ decay rate places stringent constraints on 
possible flavor physics beyond the SM.
  In Model II for example, where the most important correction
comes from the top quark and charged Higgs loop,
a $\tan\beta$-independent lower bound of about $370\;{\rm GeV}$
can be imposed on the charged Higgs mass~\cite{ciuchini}.

 This picture changes dramatically in the T2HDM where the
charm Yukawa's are greatly enhanced relative to those in Model II.
Consequently, both top and charm loops involving charged Higgs
exchange become important. Furthermore, both amplitudes
become complex due to their dependence on the new phase $\delta$.
As a result, the charged Higgs amplitude can interfere either
constructively or destructively with the SM amplitude depending 
on the phase $\delta$.
The bound on the charged Higgs mass will be dependent on both
$\tan \beta$ and $\xi$, and a relatively light Higgs can still 
be allowed.

  For the numerical estimate, we will simply neglect the effect of the 
scalar operator $\bar{c}_Rb_L\bar{s}_L c_R$ induced from Higgs exchange,
and concentrate on the effect due to the charged Higgs correction to the 
matching condition for the leading-order (LO) Wilson coefficient 
$C_7^{(0)}$ at the scale $m_W$:
\begin{eqnarray}
        \delta C^{(0)}_7(m_W)\!\! & = &\!\!\! \sum_{u=c,t}\kappa^u
                \left[ -\tan^2\beta +  
                \left(\Sigma^T V^\ast\right)_{us}\left(\tan^2\beta +1
  \right)/ m_u V_{us}^\ast \right]  \label{eq:delc7} \\
               \!\! & &\!\!\! \times\left\{B(y_u) + A(y_u)
                        \left[-1 + 
                \left(\Sigma^\dagger V\right)_{ub} \left(\cot^2\beta +
 1\right)/ m_u V_{ub} \right]/6 \right\} . \nonumber
\end{eqnarray}
\noindent In this expression $\kappa^u = \pm 1$ for $u=c,t$,
 $y_u = (m_u/m_H)^2$, and
$A$ and $B$ are the standard expressions~\cite{grinstein}.
The effective Wilson coefficient at the scale $\mu \sim {\cal O}(m_b)$
is modified according to
\begin{equation}
C_7^{(0)\rm eff} (\mu)\rightarrow C_7^{(0)\rm eff}(\mu)+
 (\alpha_s(m_W)/\alpha_s(\mu))^{16/23} \delta C_7^{(0)}(m_W)\; . 
\end{equation}
The $b\to s \gamma$ constraint on the $(m_H, \tan\beta)$ plane,
together with the constraints from $\Delta m_K$ and $\epsilon_K$,
are presented in Fig.~1 for three representative choices of the
CKM phase $\gamma$.  

  The experimental data on $B\to X_c \tau \nu$ and on other decays
do not lead to new exclusion regions 
after imposing the $b\to s \gamma$ and $K\bar{K}$ constraints.

\vskip .5 cm

\begin{figure}[ht]  
\centerline{\epsfxsize 6 truein \epsfbox{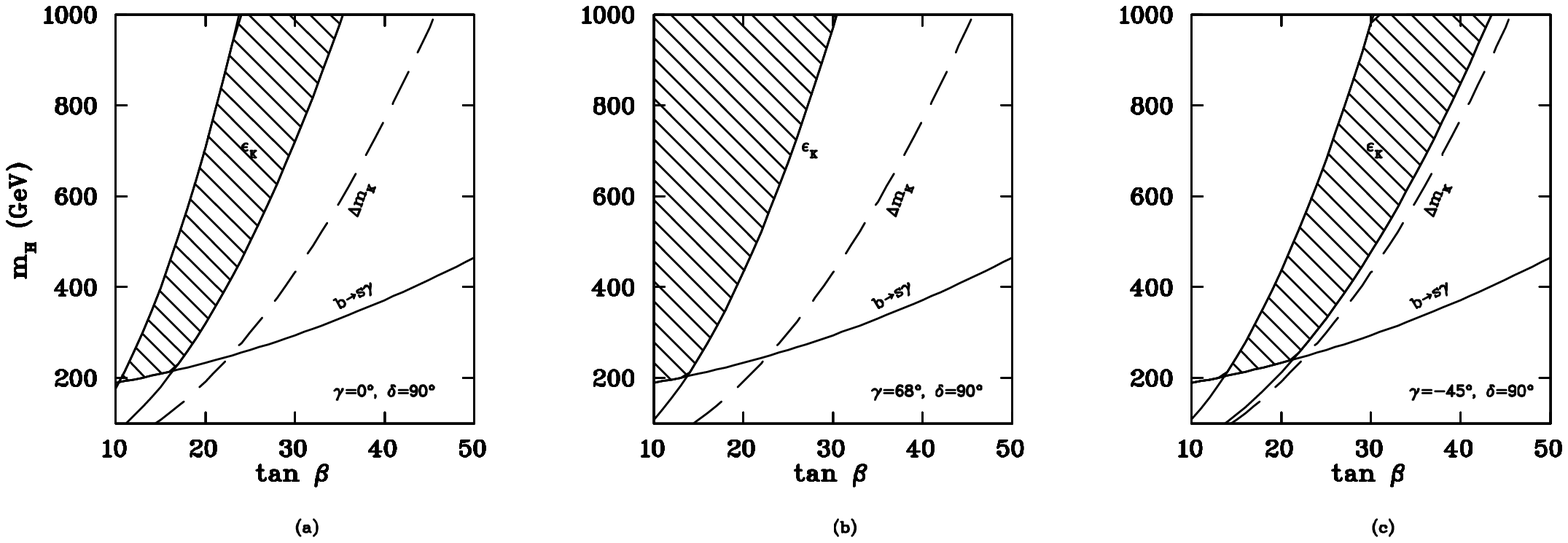}}
\vskip .2 cm
\caption[]{    
\label{fig1}      
\small Experimental constraints on the T2HDM from $b\to s \gamma$,
$\epsilon_K$ and $\Delta m_K$ for three representative choices
of the CKM phase $\gamma$ and for $\delta=90^\circ$ and $|\xi|=1$:
(a) $\gamma=0^\circ$, (b) $\gamma=68^\circ$ (SM central value),
(c) $\gamma=-45^\circ$.
   The allowed regions are shaded. } 
\end{figure}

\section{$CP$ Asymmetry in $B \to J/\psi K_S$}

  In the SM, the time-dependent $CP$ asymmetry 
$a(t)\equiv\left[\Gamma(B(t))-\Gamma(\overline{B}(t))\right]/
\left[\Gamma(B(t))+\Gamma(\overline{B}(t))\right]$ for $B\to \psi
K_S$ is free of hadronic uncertainties and is given by
$a_{\rm SM}(t) = - \sin\left( 2\beta_{\rm CKM}\right)\sin
        \left(\Delta Mt\right)$~\cite{hquinn},
where $\Delta M = M_{B_H}-M_{B_L}$ is the mass difference between
the neutral $B$ mesons. Existing experimental data indirectly constrains
$\sin(2\beta_{\rm CKM}) = .75\pm.10$ \cite{paganini} within the SM.
Note that this  $CP$ asymmetry in the SM  arises from the 
$B\bar{B}$ mixing \cite{grolon}
 and not from the decay amplitudes, as can be seen easily
from the Wolfenstein parameterization of the CKM matrix.

  In the T2HDM,  there is one additional diagram due to charged Higgs
exchange that mediates the decay $B\to \psi K_S$.
Integrating out the $W$ and $H^+$ gives us
the effective Lagrangian,
\begin{equation}
        {\cal L}_{\rm eff} \simeq -2\sqrt{2} G_F V_{cb}V_{cs}^\ast
                \left[ \overline{c}_L\gamma_\mu b_L  
                        \overline{s}_L\gamma^\mu c_L
                +2\zeta e^{i\delta}
                \overline{c}_Rb_L\overline{s}_Lc_R\right] +{\rm h.c.},
        \label{eqn:leff}
\end{equation}
\noindent where $\zeta e^{i\delta} \equiv (1/2)(V_{tb}/V_{cb})
(m_c\tan\beta/m_H)^2\xi^\ast$  with $\zeta$ taken to be real and positive.
Terms which are subdominant in the large $\tan \beta$ limit are not 
included in the above equation.
Note that the charged-Higgs-mediated  decay amplitude is complex and its
interference with the SM amplitude could significantly modify the 
$CP$ asymmetry in $B \to J/\psi K_S$ \cite{growor}. On the other hand, the 
charged Higgs has a vanishingly small effect on the 
$B\bar{B}$ mixing amplitude. 

  To proceed, we assume factorization and use Fierz rearrangement
to evaluate the hadronic matrix elements of the two four-Fermi operators.  
The total amplitudes are then obtained,
\equation 
{\cal A} \equiv {\cal A}(B\to\psi K_S) \simeq
                {\cal A}_{\rm SM}
                        \left[ 1 - \zeta e^{-i\delta}\right] 
\; , \;\;\;\;\;\;\;\;\;\;\;\;\;\;
\overline{\cal A} \equiv \overline{\cal A}(\bar{B}\to\psi K_S)
\simeq \overline{\cal A}_{\rm SM}
                        \left[ 1 - \zeta e^{i\delta}\right]  
\endequation
where the SM amplitudes satisfy 
${\cal A}_{\rm SM}=\overline{\cal A}_{\rm SM}$.
Therefore the ratio $\overline{\cal A}/{\cal A}$ gets
a phase,
\equation
\overline{\cal A}/{\cal A} =\exp(- 2 i \vartheta) \; ,
\;\;\;\;\;\;\;\;\;\;\;\;\;\;\;\;\;\;\;\;
 \tan\vartheta = \zeta\sin\delta/ (1-\zeta\cos\delta)
\endequation
Instead of measuring $\sin 2\beta_{\rm CKM}$, the 
$CP$ asymmetry nows measures $\sin 2(\beta_{\rm CKM}+ \vartheta)$

  Although the magnitude of $\vartheta$ can be at most of order $10^\circ$
after taking into account the experimental constraints,
there could  be large deviations of the $CP$ asymmetry, 
$a_{\psi K_S} \equiv \sin (2\beta_{\rm CKM} + 2 \vartheta)$,
from the SM prediction. This can be understood as follows.
The charged Higgs contribution to $\epsilon_K$ 
could be comparable in size to the SM effect,
and this basically sets free the CKM angles $\beta_{\rm CKM}$ and
$\gamma$. When $\beta_{\rm CKM}$ and $\gamma$ take negative values,
the $CP$ asymmetry in $B\to J/\psi K_S$ can be of opposite sign
relative to the SM expectation.
The CP asymmetry in the present model is illustrated in Fig.~2 for three
representative choices of $\gamma$.
As can be seen from the figure, $a_{\psi K_S}$ can take almost
any value in the T2HDM.

\begin{figure}[ht]	
\centerline{\epsfxsize 2.5 truein \epsfbox{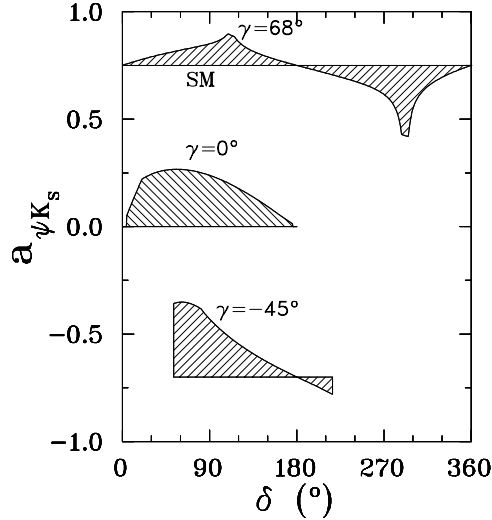}}   
\vskip .2 cm
\caption[]{
\label{fig2}
\small Amplitude of the time-dependent CP asymmetry
$a_{\psi K_S} \equiv \sin (2\beta_{\rm CKM} + 2 \vartheta)$
  versus the non-standard
CP-odd phase $\delta$ for $B\to\psi K_S$.
The top horizontal line is for the SM
assuming the best fit value ($\sin 2\beta_{\rm CKM}=0.75$) of
Ref.~\cite{paganini}.  
The shaded regions correspond to the allowed ranges of the
asymmetry in the T2HDM for three representative
choices of $\gamma$: $\gamma=68^\circ$ is the best fit of
Ref.~\cite{paganini}, $\gamma=0^\circ$ corresponds to a real CKM matrix, and
$\gamma=-45^\circ$ changes the sign of $a_{\psi K_S}$ relative to
the SM expectation.}
\end{figure}

\section{discussion}

  The analysis for $B\to J/\psi K_S$ can be straightforwardly
extended to other $B$ decay modes \cite{prep}. Of particular interest
are those decays where the SM prediction for the CP asymmetry
is zero or very small.  
These include $B_s \to J/\psi \phi$, $B_s \to J/\psi K_S$, and some others.
In the T2HDM, it is not hard to see that the $CP$ asymmetries
for these two decays are $\sim \sin 2 \vartheta$, and thus could
be of order tens of percent. Indeed, a simple 
$CP$-violating pattern can be obtained for the various $B$ decays
that is quite distinctive from the SM prediction.
 
  The complex, enhanced charm Yukawa for $H^+\bar{c}_Rd_L$
provides an interesting scenario for generating a sizable neutron
electric dipole moment (EDM) via a one loop diagram with virtual
charm and $H^+$. Two vertices of the diagram involve 
charm quarks of opposite chirality in order for the diagram to 
have an imaginary piece.  
Using the non-relativistic quark model 
relation between the neutron EDM and quark EDM, one has,
\equation
d_n \simeq \frac{4}{3} d_d 
    = \frac{\sqrt{2}G_Fm_d}{9\pi^2} 
      \frac{m_c^2 \tan^2\beta}{m^2_H}
      A |\xi| \sqrt{(1-\rho)^2 + \eta^2} \lambda^4 
      \sin (\beta_{\rm CKM} -\delta)
       \left( \ln \frac{m^2_H}{m^2_c} - \frac{3}{4} \right) \; .
\endequation
Taking $m_d$ as the current quark mass and imposing the 
experimental constraints on the model, we find that the neutron
EDM can be as large as $10^{-27} - 10^{-26} \; e\cdot {\rm cm}$,
not too far below the present experimental limit.
 It is worth pointing out that this estimate is valid even if the new
phase $\delta$ vanishes, as the $c_R-t_R$ mixing always gives 
the $H^+\bar{c}_Rd_L$ coupling a complex piece that is proportional 
to $V_{td}$.

  The top-quark 2HDM provides one example for new $CP$-violating phases
and non-standard charged Higgs Yukawa couplings which should be 
generally present in multi-Higgs doublet models when natural flavor
conservation is not enforced.  Although one need not take this particular 
model too seriously, the interesting phenomenological implications
of flavor violation in general deserve further investigation.


\begin{references}  
  

\bibitem{paganini}
P. Paganini {\it et al}., Phys. Scripta {\bf 58}, 556 (1998); 
see also S. Mele, hep-ph/9808411. 

\bibitem{nfc}
S. Glashow and S. Weinberg, Phys. Rev. D {\bf 15}, 1958 (1977).

\bibitem{nonfc}
T.P. Cheng and M. Sher, Phys. Rev. D {\bf 35}, 3484 (1987);\\
L. Hall and S. Weinberg, Phys. Rev. D {\bf 48}, R979 (1993); \\ 
Y. L. Wu and L. Wolfenstein, Phys.  Rev.  Lett.  {\bf 73}, 1762 (1994), 
and Y. L. Wu, hep-ph/9404241.

\bibitem{type3} 
For a recent analysis, see
D. Atwood, L. Reina, and A. Soni, Phys.\ Rev.\ D {\bf 55}, 3156 (1997),
and references therein.

\bibitem{hhunter}
For a review, see J. Gunion {\it et. al.}, {\em The Higgs Hunter's Guide},
Addison-Wesley Publishing Company, 1990.

\bibitem{daskao}
        A. Das and C. Kao, Phys.\ Lett.\ B {\bf 372}, 106 (1996). 

\bibitem{ksw}
K. Kiers, A. Soni, and G.-H. Wu, hep-ph/9810552, to appear in Phys. Rev. D.

\bibitem{buchalla}
For a recent review, see, for example
 G. Buchalla, A.J. Buras, and M.E. Lautenbacher,
 Rev.\ Mod.\ Phys.\ {\bf 68}, 1125 (1996). 

\bibitem{wolfen}
  L. Wolfenstein, Phys.\ Rev.\ Lett.\  {\bf 51}, 1841 (1983).  


\bibitem{ciuchini}  See, e.g.,
        M. Ciuchini, {\it et al}., Nucl. Phys. {\bf B527}, 21 (1998).  

\bibitem{grinstein}
    B. Grinstein, R. Springer, and M. Wise,
 Nucl. Phys. {\bf B339}, 269 (1990). 

\bibitem{hquinn} See, e.g., the article by H. Quinn in Review of
Particle Physics, C. Caso {\it et al}., Eur. Phys. J. C
{\bf 3}, 1 (1998).  

\bibitem{grolon}
For a recent analysis of new physics effects on $CP$ asymmetries
through $CP$ violation in $B\bar{B}$ mixing, see for example,
 M. Gronau and D. London, Phys. Rev. D {\bf 55}, 2845 (1997), and 
references therein.

\bibitem{growor}
For related work, see
Y. Grossman and M. Worah in Phys. Lett. B {\bf 395}, 241 (1997). 

\bibitem{prep}
Manuscript in preparation.

\end{references}
\end{document}